\documentclass[aps,pra,twocolumn,showpacs,balancelastpage,superscriptaddress,groupedaddress]{revtex4-1}  

\usepackage{color}
\usepackage[T1]{fontenc}
\usepackage[latin1]{inputenc}
\usepackage[english]{babel}
\usepackage[full]{complexity}
\usepackage{hyperref}
\usepackage{amsthm}
\usepackage{graphicx,subfigure}  
\usepackage{dcolumn}   
\usepackage{bm}        
\usepackage{amssymb}   
\input{Qcircuit.tex}

\renewcommand{\Qcircuit}[1][0em]{\xymatrix @*=<#1>}

\newcommand{\hide}[1]{}

\theoremstyle{definition}
\newtheorem{thm}{Theorem}
\newtheorem{lem}[thm]{Lemma}

\newcommand{\SSS}{${\mathcal S}$}

\begin{document}

\title{Linear-Depth Quantum Circuits for $n$-qubit Toffoli gates with no Ancilla}

\author{Mehdi Saeedi}%
\thanks{Corresponding author: msaeedi@usc.edu}
\author{Massoud Pedram}%
\affiliation{Department of Electrical Engineering, University of Southern California, Los Angeles, CA 90089-2562}

\begin{abstract}
We design a circuit structure with linear depth to implement an $n$-qubit Toffoli gate. The proposed construction uses a quadratic-size circuit consists of elementary 2-qubit controlled-rotation gates around the $x$ axis and uses no ancilla qubit. Circuit depth remains linear in quantum technologies with finite-distance interactions between qubits. The suggested construction is related to the long-standing construction by Barenco et al. (Phys. Rev. A, 52: 3457-3467, 1995) \cite{Barenco95}, which uses a quadratic-size, quadratic-depth quantum circuit for an $n$-qubit Toffoli gate.
\end{abstract}

\maketitle

\section{Introduction}
Practical implementation of multi-qubit quantum gates in quest of a scalable quantum computing system is essential. In particular, an $n$-qubit Toffoli gate plays a key role in established quantum algorithms. Examples include compiled circuits for modular multiplication and exponentiation in Shor's number-factoring algorithm \cite{VanMeter,MarkovQIC2012,Markov2013} and quantum error correction codes \cite{NeilsenChuang}. For $n=3$, the entangling Toffoli gate, which flips `target' state conditioned on its two `controls', is universal in reversible Boolean logic, see \cite{SaeediM2011}. Additionally, with an appropriate single-qubit gate, the 3-qubit Toffoli gate constructs a universal gate set for quantum computing \cite{Shi:2003}. In the recent years, several protocols have been proposed to realize the 3-qubit Toffoli gate and its variants in different physical quantum technologies, e.g., with superconducting qubits \cite{Fedorov12,Stojanovic12}, trapped ions \cite{Monz09,Borrelli11}, optical elements \cite{Ralph07,Lanyon08}, and cavity quantum electrodynamics \cite{Shao07}.

A common approach to implement a highly conditional gate is to apply \emph{decomposition} which breakdowns the gate into `elementary' gates with at most one control \cite{Mikko,Shende06,SaeediQIC11}. For an $n$-qubit Toffoli gate, this path results in quadratic-size, quadratic-depth quantum circuits with no ancilla \cite[Corollary 7.6]{Barenco95}.
For the 3-qubit Toffoli gate, the simplest known decomposition requires five 2-qubit gates \cite[Lemma 6.1]{Barenco95}, or exactly six CNOTs \cite{Shende09} and several one-qubit gates. To avoid applying a long, at least quadratic-length, sequence of single- and 2-qubit gates, several methods have been proposed to directly realize multi-qubit gates with trapped ions \cite{Wang01,Ivanov11}, neutral atoms \cite{Duan05}, or superconducting qubits \cite{Yang05}.

To streamline the realization of Toffoli gates conditioned on many qubits, which can speed-up the progress towards scalable quantum computation, both theoretical and experimental attempts are extremely important. In this paper, we propose a theoretical approach to decompose $n$-qubit Toffoli gates into 2-qubit gates in quadratic size, but linear depth, without using additional ancilla qubit. For this purpose, we change the usual computational basis states $\ket{0}$ and $\ket{1}$ and propose a construction which exploits quantum rotation gates conditioned on one qubit. The proposed construction is related to the synthesis framework we suggested in \cite{Abdollahi}.

The rest of this paper is organized as follows. The proposed circuit structure is introduced in Section \ref{sec:ours}. Circuit depth is analyzed in Section \ref{sec:depth} for quantum computing systems with arbitrary-length and finite-length interaction distance between qubits. We compare the proposed structure with prior constructions in Section \ref{sec:comp}. Section \ref{sec:conc} concludes the paper with further discussion.

\section{Circuit Structure} \label{sec:ours}
The choice of basis states in quantum computing is not unique and any two orthogonal unit vectors can be used in a 2-particle quantum computing system to serve as the computational basis states. Working with rotation gates $R_x(\pi)$ around the $x$ axis, we keep $\hat 0 =\ket{0}$, but change the other vector to $\hat 1=R_x(\pi)\ket{0}=  \left[ {\begin{array}{*{20}c} 0 & -i  \\ \end{array}} \right]^T$. Accordingly, $R_x(\pi)$ works as a NOT gate which transforms $\hat 0$ to $\hat 1$ and vice versa. Adding one and two conditions for $R_x(\pi)$ constructs analogous versions of the conventional 2-qubit CNOT and 3-qubit Toffoli gates. Accordingly, an $n$-qubit Toffoli gate is a $\pi$-rotation gate around the $x$ axis with $n-1$ conditionals. In circuit diagrams throughout the paper, $k$ consecutive gates with the same control lines are shown as a single gate with one control and $k$ targets. Furthermore, $C^iR_x(\theta)$ is a $\theta$-rotation gate around the $x$ axis with $i$ controls, $a R_x(\theta) b$ is a $\theta$-rotation gate on $b$ conditioned on $a$, and $a_1R_x(\theta_1) (a_2 R_x(\theta_2) a_3)=a_2 R_x(\theta_2) (a_1 R_x(\theta_1) a_3)$.

Figure \ref{Fig:CTGates} shows a possible decomposition for a 3-qubit Toffoli gate. In this figure, if at least one of the first two qubits is $\hat{0}$, then the circuit applies either an identity $I$ gate or $R_x(\frac{\pi}{2}-\frac{\pi}{2})=I$ gate to the target qubit. Otherwise, $R_x(\frac{\pi}{2}+\frac{\pi}{2})$ is applied which is a NOT gate. Lemma \ref{lem:QIC} provides a hierarchical structure for an $n$-qubit Toffoli gate after applying an automatic synthesis framework in \cite{Abdollahi}. In the following, we first prove the lemma and then use it in our construction.

\begin{figure}[tb]
\scalebox{0.8}{
\input{figures/CNOT-Toffoli}
}
\centering
\caption{\label{Fig:CTGates} The 3-qubit Toffoli gate and its decomposition into 2-qubit controlled-rotation gates. Two consecutive gates with controls on $a$ are
shown as a single gate with one control and two targets on $b$ and $c$.}
\end{figure}

\begin{lem} \cite[Section 6]{Abdollahi} \label{lem:QIC}
An $n$-qubit Toffoli gate can be implemented as (\ref{eq:QIC}).
\end{lem}

\small
\small
\begin{align}\label{eq:QIC}
C^{n-1}R_x(\pi) &= C^{n-2}R_x(\pi) R_x(-\pi/2)  \nonumber \\
     &\qquad \,\,\,\,\,\, [C^{n-3}R_x(\pi) R_x(-\pi/4)  \nonumber \\
       &\qquad \,\,\,\,\,\,\,\,\,\,\,\,\,\,\,\,\,\,     [\cdots  \nonumber \\
       &\qquad \,\,\,\,\,\,\,\,\,\,\,\,\,\,\,\,\,\,\,\,\,\,\,\,     [C^{n-(n-1)}R_x(\pi) R_x(-\pi/2^{n-2}) \nonumber \\
        &\qquad         [a_1 R_x(\pi/2^{n-2})(a_2 R_x(\pi/2^{n-2}) \nonumber \\
        &\qquad         (a_3 R_x(\pi/2^{n-3})(\cdots (a_{n-1} R_x(\pi/2) a_{n} ) \cdots )] \cdots ]\nonumber \\
\end{align}
\normalsize

\begin{figure*}[tb]
\scalebox{0.8}{
\input{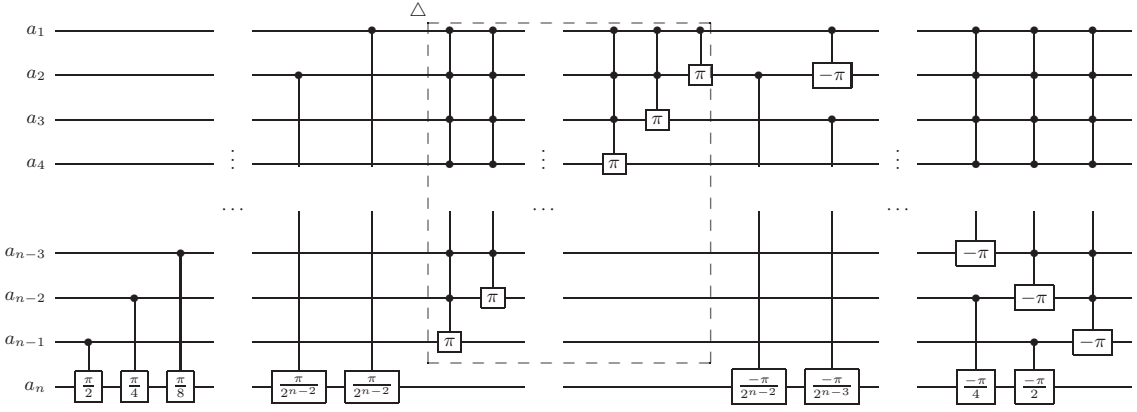}
}
\centering
\caption{\label{Fig:TheoremQIC} An $n$-qubit Toffoli gate can be implemented using 2-, 3-, $\cdots$, $(n-1)$-qubit Toffoli gates based on (\ref{eq:QIC}).}
\end{figure*}

{\bf Proof.} Equation (\ref{eq:QIC}) is illustrated in Figure \ref{Fig:TheoremQIC} with including conditional $-\pi$-rotation gates to restore control qubits. Consider the subcircuit $\triangle$ shown in the figure. Focusing on $\triangle$, assume $\triangle$ input qubits are $a_i$ and $\triangle$ output qubits are $b_i$ for $ 1 \leq i \leq n-1$. Additionally, assume that $a_k$ is the first qubit (starting from $k=1$) with value $\hat{0}$. After applying $\triangle$, we have $b_1=a_1$, $b_i=0$ for $2 \leq i \leq k-1$, $b_k=1$, and $b_i=a_i$ for $k+1 \leq i \leq n$.

Now, consider the complete circuit in Figure \ref{Fig:TheoremQIC}. The case $a_1=\hat{0}$ is trivial because gates in $\triangle$ are disabled, the gate with control qubit $a_1$ and target qubit $a_n$ is deactivated, and other applied gates cancel out the effects of each other. Therefore, we assume $a_1=\hat{1}$. Note that before applying $\triangle$, each controlled-rotation gate with control qubit $a_i$ for $2 \leq i \leq n-1$ applies $\pi/2^{n-i}$ to qubit $a_n$. Similarly, after applying $\triangle$, each controlled-rotation gate with control qubit $a_i$ for $2 \leq i \leq n-1$ applies $-\pi/2^{n-i}$ to qubit $a_n$.

If $a_k$ (starting from $k=1) $ is the first qubit with value $\hat{0}$, then conditional rotation gates with controls $a_1, a_2, \cdots, a_{k-1}$ are activated and a $\theta_1$-rotation gate with $\theta_1=\frac{\pi}{2^{n-2}}+\frac{\pi}{2^{n-2}}+\frac{\pi}{2^{n-3}}+\cdots+\frac{\pi}{2^{n-k+1}}$ is applied to the target qubit. However, after applying $\triangle$ a $\theta_2$-rotation gate with $\theta_2=\frac{-\pi}{2^{n-k}}$ is applied which removes the effect of $\theta_1$ given $\theta_1=-\theta_2$. Additionally, each gate with control qubit $a_i$ for $k+1 \leq i < n-1$ after $\triangle$ removes the effect of its corresponding gate before $\triangle$.

Finally, if $a_i=\hat{1}$ for all $1 \leq i \leq n-1$, then all gates before $\triangle$ are enabled and all gates after $\triangle$ are disabled and a $\theta$-rotation gate with $\theta=\frac{\pi}{2^{n-2}}+\frac{\pi}{2^{n-2}}+\frac{\pi}{2^{n-3}}+\cdots+\frac{\pi}{2^2}+\frac{\pi}{2}=\pi$ is applied to the target qubit $a_n$. \qed

\begin{figure*}[tb]
\scalebox{0.8}{
\input{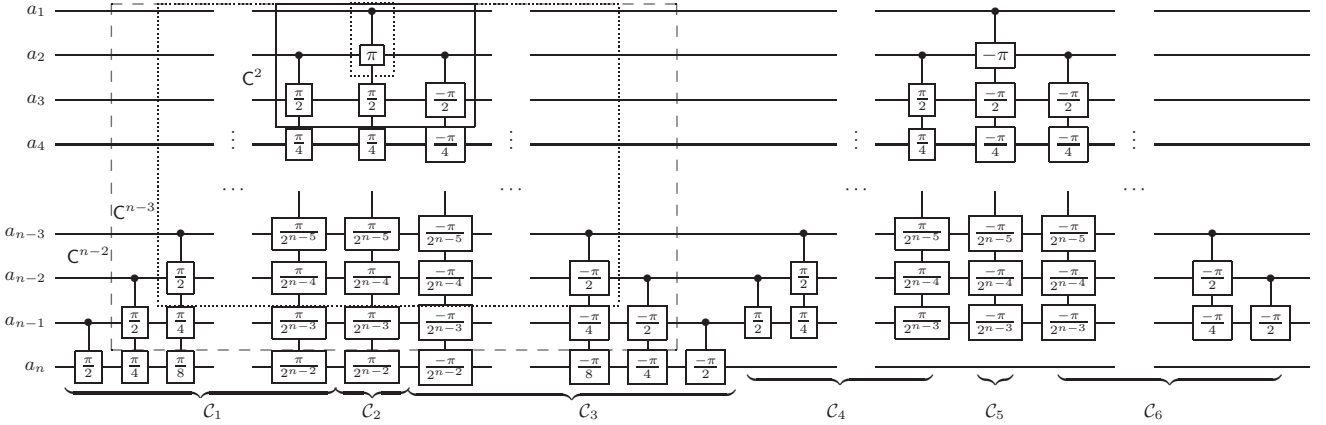}
}
\centering
\caption{\label{Fig:Theorem} Circuit structure for an $n$-qubit Toffoli gate. Excluding gates that should be applied to restore control qubits, different boxes represent $(n-1)$-qubit Toffoli, $(n-2)$-qubit Toffoli, $\cdots$, 3-qubit Toffoli, and 2-qubit CNOT gates. These gates are specified by $C^i$, for $C^iR_x(\pi)$, in the figure. The proposed construction is divided into six parts $\mathcal{C}_1, \mathcal{C}_2, \cdots, \mathcal{C}_6$.}
\end{figure*}

\begin{thm} \label{thm:strucutre}
An $n$-qubit Toffoli gate with controls $a_1,a_2,\cdots,a_{n-1}$ and target $a_{n}$ can be implemented by a network of the form given in Figure \ref{Fig:Theorem} where all gates are conditional $\theta$-rotation gates around the $x$ axis.
\end{thm}
{\bf Proof.} To prove, we restructure the circuit shown in Figure \ref{Fig:Theorem} as illustrated in Figure \ref{Fig:TheoremQIC}. To verify, note that gates in the first (top) $n-1$ lines construct an $(n-1)$-qubit Toffoli gate, gates in the first $n-2$ lines construct an $(n-2)$-qubit Toffoli gate, $\cdots$, gates in the first 3 qubit constructs a 3-qubit Toffoli, and finally gate in the first 2 qubits is a CNOT, these gates are specified in Figure \ref{Fig:Theorem} too. Based on Lemma \ref{lem:QIC}, the circuit shown in Figure \ref{Fig:Theorem} implements an $n$-qubit Toffoli gate. \qed

\begin{figure}[tb]
\scalebox{0.8}{
\input{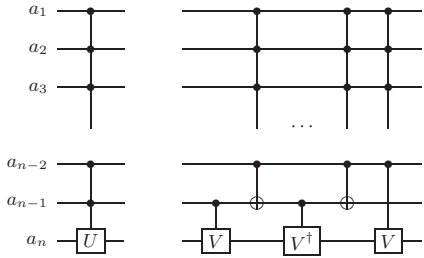}
}
\centering
\caption{\label{Fig:Barenco} Circuit for an $n$-qubit Toffoli gate in \cite[Lemma 7.5]{Barenco95} where $V^2=U$ and $U$ is a NOT gate.
The last gate can be decomposed by recursively applying the decomposition. }
\end{figure}

Figure \ref{Fig:T5-1} shows the proposed construction for a 5-qubit Toffoli gate. In Figure \ref{Fig:T5-2}, the construction is restructured differently to better reflect the hierarchial structure in \cite{Abdollahi}. Comparing the decomposition in Figure \ref{Fig:T5-2} with the conventional method in \cite{Barenco95}, see Figure \ref{Fig:Barenco}, reveals the main differences between two methods.
To count the number of 2-qubit gates in the proposed construction, note that there are $2\Sigma_{i=1}^{i=n-2}i+n-1$ gates to construct the transformation on the target line, and $2\Sigma_{i=1}^{i=n-3}i+n-2$ gates to restore control lines to their original values. Therefore, the total number of 2-qubit gates in the proposed construction is $2n^2-6n+5$ or $2n^2+O(n)$.

\section{Depth Analysis} \label{sec:depth}
\begin{figure}[t]
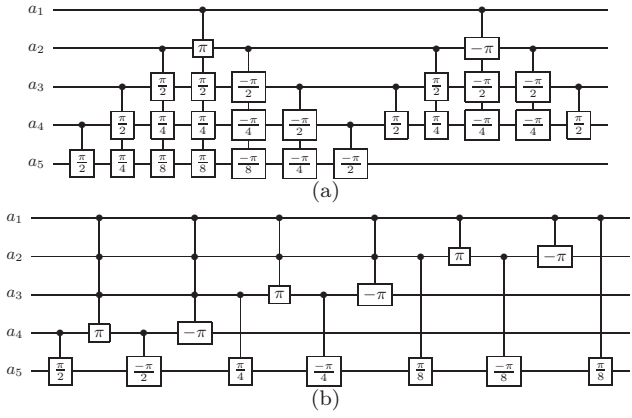

    \centering
    \subfigure[\label{Fig:T5-1} ]{
        \scalebox{0.7}{
        \input{figures/T5-1}
        }
    }
    \subfigure[\label{Fig:T5-2} ]{
        \scalebox{0.7}{
        \input{figures/T5-2}
        }
    }
    \caption{\label{Fig:T5} Circuit structure for a $5$-qubit Toffoli gate. Circuit in (a) is the proposed structure. This circuit is restructured in (b) to use circuits for 4, 3, and 2-qubit Toffoli and CNOT gates. Note that direct decomposition of the gates in (b) results in many redundant gates.
    }
\end{figure}

\begin{figure}[t]
\scalebox{0.8}{
\input{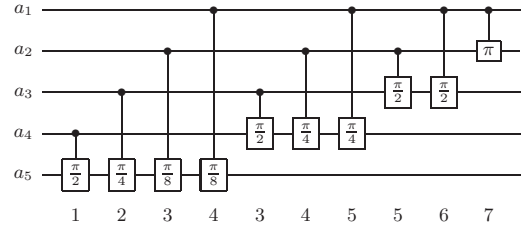}
}
\centering
\caption{\label{Fig:Parallel} A part of the circuit shown in Figure \ref{Fig:T5-1} restructured to show parallel circuits. Numbers are the time slots that gates can be executed. }
\end{figure}

In this section, we show that in spite of the quadratic size of the proposed structure for an $n$-qubit Toffoli gate (no ancilla), circuit depth is linear. 
In order to consider depth, we restructure the construction shown in Figure \ref{Fig:Theorem}. In particular, we change the structure to have gates with common targets (vs. common controls in Figure \ref{Fig:Theorem}) in sequence. Additionally, we divide the circuit in Figure \ref{Fig:Theorem} into 6 parts, namely $\mathcal{C}_1, \mathcal{C}_2, \cdots, \mathcal{C}_6$ as shown in the figure. To evaluate circuit depth, we focus on $\mathcal{C}_1$. The result can be extended to the whole circuit. Figure \ref{Fig:Parallel} illustrates $\mathcal{C}_1$ in Figure \ref{Fig:T5-1} with time steps for each gate.

\begin{thm}
The proposed structure for an $n$-qubit Toffoli gate can be implemented by a linear-depth circuit.
\end{thm}

{\bf Proof.} Restructuring the circuit structure in Figure \ref{Fig:Theorem} to have gates with common targets in sequence, one can verify that in $\mathcal{C}_1$+$\mathcal{C}_2$ there are $n-1$ gates with targets on qubit $n$, $n-2$ gates with targets on qubit $n-1$, $\cdots$, one gate with target on qubit 2. Assign time steps $1, 2, \cdots, n-1$ to $n-1$ gates with targets on qubit $n$. Next, consider the $n-2$ gates with targets on qubit $n-1$. Among these gates, $n-3$ gates can be executed in parallel with the gates with targets on qubit $n$. Precisely, gates with targets on qubit $n-2$ can be executed in time steps $3, 4, \cdots, n-1, n$. Similarly, the next $n-4$ gates can be executed in time steps $5, 6, \cdots, n+1$. Following this path results in $2n-3$ time steps for $\mathcal{C}_1$+$\mathcal{C}_2$. Likewise, $\mathcal{C}_3$ can be parallelized to depth $2n-5$, $\mathcal{C}_4+\mathcal{C}_5$ can be parallelized to depth $2n-5$, and finally $\mathcal{C}_6$ can be parallelized to depth $2n-7$. Altogether, circuit depth for an $n$-qubit Toffoli gate in the proposed construction is $8n-20$. \qed

While circuit depth in the proposed construction is linear, our construction includes many long-distance 2-qubit gates. In general, restricting interactions to only linear dimension (1D) results in $O(n)$ overhead. However, circuit depth in the proposed construction remains linear even in very restrictive quantum architectures with possible interactions in a line. Assume a SWAP gate between qubits $a_1$ and $a_2$ is represented by \SSS$(a_1,a_2)$. We use the term `local' for gates that use \emph{neighbor} qubits in a given architecture.

\begin{thm} \label{thm:swap}
Circuit depth for an $n$-qubit Toffoli in the proposed construction is linear in architectures with finite-distance interactions between qubits.
\end{thm}

{\bf Proof.} To prove, we consider 1D architectures. One can execute a 1D quantum circuit on architectures with interactions in a higher dimension. Working with $\mathcal{C}_1+\mathcal{C}_2$, consider a chain of $n-1$ serial SWAP gates \SSS$(a_n,a_{n-1})$, \SSS$(a_{n-1},a_{n-2})$, \SSS$(a_{n-2},a_{n-3})$, $\cdots$, \SSS$(a_{2},a_{1})$ in sequence. For an initial qubit ordering $1,2,\cdots, n$, the resulting ordering is $n, 1, 2, \cdots, n-1$ (i.e., a 1-bit rotation). Immediately after each SWAP gate, one can apply a local controlled-rotation gate with target on qubit $n$. Now, apply a chain of $n-2$ SWAP gates \SSS$(a_n,a_{n-1})$, \SSS$(a_{n-1},a_{n-2})$, \SSS$(a_{n-2},a_{n-3})$, $\cdots$, \SSS$(a_{3},a_{2})$ in sequence. Among these $n-2$ gates, $n-3$ gates can be executed in parallel with the previous gates. After the second SWAP chain, the resulting qubit ordering is $n, n-1, 1, 2, \cdots, n-2$, i.e., a 2-bit rotation. Accordingly, we can apply $n-2$ local controlled-rotation gates with targets on $n-1$. Following this path results in $2n-3$ time steps for SWAP gates, and $2n-3$ time steps for controlled-rotation gates, $4n-6$ 2-qubit time steps in total. Circuit size is increased by $2n-3$ for SWAPs. The final qubit ordering is $n, n-1, n-2, \cdots, 2, 1$.

To construct a local circuit for $\mathcal{C}_3$ starting from qubit ordering $n, n-1, n-2, \cdots, 2, 1$, we can apply the same structure discussed. It leads to depth $4n-10$ for $\mathcal{C}_3$. The resulting qubit ordering is $2,3,\cdots, n-1, n, 1$. At this time, applying the next $\mathcal{C}_4+\mathcal{C}_5$ circuit is tricky because qubit ordering has been changed from the initial one $1,2,\cdots,n-1, n$. Actually, the first qubit is far from other qubits $2, 3, \cdots$. For this case, we apply a linear-depth circuit with depth $n+5$, and size $4n-6$ \cite[Theorem 4.1]{Kutin07} to restore the ordering $1,2,\cdots, n-1, n$. Accordingly, $\mathcal{C}_4+\mathcal{C}_5$, and $\mathcal{C}_6$ can be implemented in depth $4n-10$ and $4n-14$, respectively. We recover the final qubit ordering to the initial ordering $1,2,\cdots,n-1, n$ with another linear-depth circuit.

Altogether, circuit depth for an $n$-qubit Toffoli gate with only 1D interactions can be calculated as $18n-31$. Circuit size remains $2n^2+O(n)$.  \qed

\begin{figure}[t]
\scalebox{0.8}{
\input{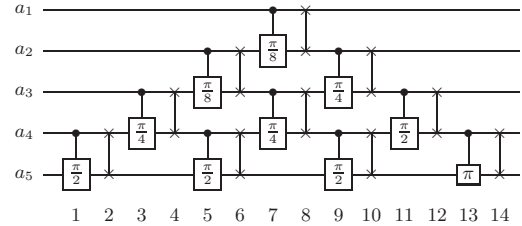}
}
\centering
\caption{\label{Fig:SWAP} Circuit in Figure \ref{Fig:Parallel} with only local gates. Numbers are time slots that gates can be executed. }
\end{figure}

In summary, circuit depth in the proposed structure is only increased by a constant factor, e.g., 2.25 in 1D architectures. Figure \ref{Fig:SWAP} illustrates the circuit in Figure \ref{Fig:Parallel} with only local gates.

\section{Comparison with prior art} \label{sec:comp}
The current widely-used decomposition \cite[Corollary 7.6]{Barenco95} for an $n$-qubit Toffoli gate uses a quadratic-size construction with staircase structure where target of gate $i$ depends on a control of gate $i-1$. This results in a quadratic depth. The decomposition is illustrated in Figure \ref{Fig:Barenco}. In this figure, $U$ is a NOT gate which results in $V$ and $V^\dagger$ where $V^2=U$. The resulting multiple-control Toffoli gates have linear cost $48n+O(1)$ in \cite{Barenco95} due to the availability of one ancilla qubit. The last gate can be decomposed by recursively applying the decomposition shown in Figure \ref{Fig:Barenco} using $U=\sqrt{\rm{NOT}}$. Following this path results in controlled-$i$th-root-of-NOT gates for $i=2^1, 2^2, \cdots, 2^{n-1}$. Circuit size and depth are $48n^2+O(n)$ 2-qubit gates.

The optimizations in \cite{MaslovTCAD} improve the linear-cost implementation of a multiple-control Toffoli gates with one ancilla from $48n+O(1)$ to $24n+O(1)$. Circuit depth remains quadratic, precisely $24n^2+O(n)$. The method in \cite[Section 6]{Abdollahi} benefits from a recursive construction with quadratic-depth $2n^2+O(n)$. As discussed in Section \ref{sec:ours} and Section \ref{sec:depth}, our circuit size and depth are quadratic and linear, respectively. All methods uses gates with similar complexity levels for physical realization.

In the proposed structure we assumed no ancilla qubit is available to facilitate circuit construction. If at least one ancilla exists, prior circuit structures in \cite[Lemma 7.2]{Barenco95} and \cite[Lemma 7.3]{Barenco95}, and the extended versions \cite{MaslovTCAD}, use linear-size circuits. When 1 and $n-3$ ancillae are available, we can apply the same circuit structures in \cite[Lemma 7.2]{Barenco95} and \cite[Lemma 7.3]{Barenco95}. Precisely, after applying various optimizations in \cite{MaslovTCAD}, we can construct circuits with sizes $24n - 88$, and $12n - 34$ if one and $n-3$ ancillae are available --- note that Peres gate has cost 4 in the proposed construction as in \cite{MaslovTCAD}. Reusing optimizations in \cite{MaslovTCAD} in the proposed circuit structure is straightforward.

\section{Conclusion } \label{sec:conc}
We proposed a linear-depth quadratic-size quantum circuit with controlled-rotation gates around the $x$ axis with no ancilla qubit to implement an $n$-qubit Toffoli gate. Restricting qubit interactions in finite length affects circuit depth and size by a constant factor. 

The proposed structure may or may not be a physically realizable construction in a particular quantum computing technology. The physical implementations of quantum gates are imperfect due to various reasons including decoherence and error in experimental setups. In the proposed circuit structure, we used $\theta$-rotation gates around the $x$ axis for $\theta=\frac{\pi}{2^{k}}$ and $1 \leq k \leq {n-2}$. Obviously, $\frac{\pi}{2^{n-2}}$ can be very small for large $n$ values, which makes its physical implementation complicated. Small rotation angles may be ignored in specific applications, as done for approximate quantum Fourier transform \cite{BarencoAQFT}. In particular, restricting $k \leq \lceil {\log_2{n}}\rceil$ results in $\epsilon \approx \frac{\pi}{n}$ error.

Since conditional Toffoli gates are building blocks for various quantum algorithms, in-depth characterization and understanding of their operations and imperfections possibly based on quantum tomography \cite{Riebe} can be very useful. Recently, a multi-qubit phase gate with one control qubit simultaneously controlling $n$ target qubits was implemented using superconducting qubits \cite{Yang}. Since we extensively benefit from such gates in the proposed construction, applying the method in \cite{Yang} to physically realize conditional Toffoli gates based on the method presented in this paper, e.g. the small circuit for a 4-qubit Toffoli gate, may be useful.

Finally, while we use $\hat{0}$ and $\hat{1}$ for computational basis states, we can also use $\ket{0}$ and $\ket{1}$. To achieve this, one can transform $\ket{0}, \ket{1}$ to $\hat{0}, \hat{1}$ by applying $n$ single-qubit gates with the same matrix $M$ to all qubits. This should be followed by the proposed construction. Final quantum state can be restored from $\hat{0}, \hat{1}$ to $\ket{0}, \ket{1}$ by applying $M^{\dagger}$.

$$M = \left[ {\begin{array}{*{20}c} 1 & 0  \\ 0 & {  i}  \\ \end{array}} \right],
M^{\dagger} = \left[ {\begin{array}{*{20}c} 1 & 0   \\ 0 & -i  \\ \end{array}} \right]$$

As a side note, restricting to have only one type of 2-qubit gate can increase circuit depth and size by a constant factor given each 2-qubit gate can be implemented by a constant-size circuit \cite{Barenco95}.


\begin{thebibliography}{28}%
\makeatletter
\providecommand \@ifxundefined [1]{%
 \@ifx{#1\undefined}
}%
\providecommand \@ifnum [1]{%
 \ifnum #1\expandafter \@firstoftwo
 \else \expandafter \@secondoftwo
 \fi
}%
\providecommand \@ifx [1]{%
 \ifx #1\expandafter \@firstoftwo
 \else \expandafter \@secondoftwo
 \fi
}%
\providecommand \natexlab [1]{#1}%
\providecommand \enquote  [1]{``#1''}%
\providecommand \bibnamefont  [1]{#1}%
\providecommand \bibfnamefont [1]{#1}%
\providecommand \citenamefont [1]{#1}%
\providecommand \href@noop [0]{\@secondoftwo}%
\providecommand \href [0]{\begingroup \@sanitize@url \@href}%
\providecommand \@href[1]{\@@startlink{#1}\@@href}%
\providecommand \@@href[1]{\endgroup#1\@@endlink}%
\providecommand \@sanitize@url [0]{\catcode `\\12\catcode `\$12\catcode
  `\&12\catcode `\#12\catcode `\^12\catcode `\_12\catcode `\%12\relax}%
\providecommand \@@startlink[1]{}%
\providecommand \@@endlink[0]{}%
\providecommand \url  [0]{\begingroup\@sanitize@url \@url }%
\providecommand \@url [1]{\endgroup\@href {#1}{\urlprefix }}%
\providecommand \urlprefix  [0]{URL }%
\providecommand \Eprint [0]{\href }%
\providecommand \doibase [0]{http://dx.doi.org/}%
\providecommand \selectlanguage [0]{\@gobble}%
\providecommand \bibinfo  [0]{\@secondoftwo}%
\providecommand \bibfield  [0]{\@secondoftwo}%
\providecommand \translation [1]{[#1]}%
\providecommand \BibitemOpen [0]{}%
\providecommand \bibitemStop [0]{}%
\providecommand \bibitemNoStop [0]{.\EOS\space}%
\providecommand \EOS [0]{\spacefactor3000\relax}%
\providecommand \BibitemShut  [1]{\csname bibitem#1\endcsname}%
\let\auto@bib@innerbib\@empty
\bibitem [{\citenamefont {Barenco}\ \emph {et~al.}(1995)\citenamefont
  {Barenco}, \citenamefont {Bennett}, \citenamefont {Cleve}, \citenamefont
  {DiVincenzo}, \citenamefont {Margolus}, \citenamefont {Shor}, \citenamefont
  {Sleator}, \citenamefont {Smolin},\ and\ \citenamefont
  {Weinfurter}}]{Barenco95}%
  \BibitemOpen
  \bibfield  {author} {\bibinfo {author} {\bibfnamefont {A.}~\bibnamefont
  {Barenco}}, \bibinfo {author} {\bibfnamefont {C.~H.}\ \bibnamefont
  {Bennett}}, \bibinfo {author} {\bibfnamefont {R.}~\bibnamefont {Cleve}},
  \bibinfo {author} {\bibfnamefont {D.~P.}\ \bibnamefont {DiVincenzo}},
  \bibinfo {author} {\bibfnamefont {N.}~\bibnamefont {Margolus}}, \bibinfo
  {author} {\bibfnamefont {P.}~\bibnamefont {Shor}}, \bibinfo {author}
  {\bibfnamefont {T.}~\bibnamefont {Sleator}}, \bibinfo {author} {\bibfnamefont
  {J.~A.}\ \bibnamefont {Smolin}}, \ and\ \bibinfo {author} {\bibfnamefont
  {H.}~\bibnamefont {Weinfurter}},\ }\href {\doibase 10.1103/PhysRevA.52.3457}
  {\bibfield  {journal} {\bibinfo  {journal} {Phys. Rev. A}\ }\textbf {\bibinfo
  {volume} {52}},\ \bibinfo {pages} {3457} (\bibinfo {year}
  {1995})}\BibitemShut {NoStop}%
\bibitem [{\citenamefont {Van~Meter}\ and\ \citenamefont
  {Itoh}(2005)}]{VanMeter}%
  \BibitemOpen
  \bibfield  {author} {\bibinfo {author} {\bibfnamefont {R.}~\bibnamefont
  {Van~Meter}}\ and\ \bibinfo {author} {\bibfnamefont {K.~M.}\ \bibnamefont
  {Itoh}},\ }\href {\doibase 10.1103/PhysRevA.71.052320} {\bibfield  {journal}
  {\bibinfo  {journal} {Phys. Rev. A}\ }\textbf {\bibinfo {volume} {71}},\
  \bibinfo {pages} {052320} (\bibinfo {year} {2005})}\BibitemShut {NoStop}%
\bibitem [{\citenamefont {Markov}\ and\ \citenamefont
  {Saeedi}(2012)}]{MarkovQIC2012}%
  \BibitemOpen
  \bibfield  {author} {\bibinfo {author} {\bibfnamefont {I.~L.}\ \bibnamefont
  {Markov}}\ and\ \bibinfo {author} {\bibfnamefont {M.}~\bibnamefont
  {Saeedi}},\ }\href@noop {} {\bibfield  {journal} {\bibinfo  {journal} {Quant.
  Inf. Comput.}\ }\textbf {\bibinfo {volume} {12}},\ \bibinfo {pages} {361}
  (\bibinfo {year} {2012})},\ \Eprint {http://arxiv.org/abs/arXiv:1202.6614}
  {arXiv:1202.6614} \BibitemShut {NoStop}%
\bibitem [{\citenamefont {Markov}\ and\ \citenamefont
  {Saeedi}(2013)}]{Markov2013}%
  \BibitemOpen
  \bibfield  {author} {\bibinfo {author} {\bibfnamefont {I.~L.}\ \bibnamefont
  {Markov}}\ and\ \bibinfo {author} {\bibfnamefont {M.}~\bibnamefont
  {Saeedi}},\ }\href {\doibase 10.1103/PhysRevA.87.012310} {\bibfield
  {journal} {\bibinfo  {journal} {Phys. Rev. A}\ }\textbf {\bibinfo {volume}
  {87}},\ \bibinfo {pages} {012310} (\bibinfo {year} {2013})}\BibitemShut
  {NoStop}%
\bibitem [{\citenamefont {Nielsen}\ and\ \citenamefont
  {Chuang}(2000)}]{NeilsenChuang}%
  \BibitemOpen
  \bibfield  {author} {\bibinfo {author} {\bibfnamefont {M.~A.}\ \bibnamefont
  {Nielsen}}\ and\ \bibinfo {author} {\bibfnamefont {I.~L.}\ \bibnamefont
  {Chuang}},\ }\href@noop {} {\emph {\bibinfo {title} {Quantum Computation and
  Quantum Information}}}\ (\bibinfo  {publisher} {Cambridge University Press},\
  \bibinfo {year} {2000})\BibitemShut {NoStop}%
\bibitem [{\citenamefont {Saeedi}\ and\ \citenamefont
  {Markov}(2013)}]{SaeediM2011}%
  \BibitemOpen
  \bibfield  {author} {\bibinfo {author} {\bibfnamefont {M.}~\bibnamefont
  {Saeedi}}\ and\ \bibinfo {author} {\bibfnamefont {I.~L.}\ \bibnamefont
  {Markov}},\ }\href@noop {} {\bibfield  {journal} {\bibinfo  {journal} {ACM
  Computing Surveys}\ }\textbf {\bibinfo {volume} {45}} (\bibinfo {year}
  {2013})},\ \Eprint {http://arxiv.org/abs/arXiv:1110.2574} {arXiv:1110.2574}
  \BibitemShut {NoStop}%
\bibitem [{\citenamefont {Shi}(2003)}]{Shi:2003}%
  \BibitemOpen
  \bibfield  {author} {\bibinfo {author} {\bibfnamefont {Y.}~\bibnamefont
  {Shi}},\ }\href@noop {} {\bibfield  {journal} {\bibinfo  {journal} {Quant.
  Info. Comput.}\ }\textbf {\bibinfo {volume} {3}},\ \bibinfo {pages} {84}
  (\bibinfo {year} {2003})},\ \Eprint
  {http://arxiv.org/abs/arXiv:quant-ph/0205115} {arXiv:quant-ph/0205115}
  \BibitemShut {NoStop}%
\bibitem [{\citenamefont {Fedorov}\ \emph {et~al.}(2012)\citenamefont
  {Fedorov}, \citenamefont {Steffen}, \citenamefont {Baur}, \citenamefont
  {da~Silva},\ and\ \citenamefont {Wallraff}}]{Fedorov12}%
  \BibitemOpen
  \bibfield  {author} {\bibinfo {author} {\bibfnamefont {A.}~\bibnamefont
  {Fedorov}}, \bibinfo {author} {\bibfnamefont {L.}~\bibnamefont {Steffen}},
  \bibinfo {author} {\bibfnamefont {M.}~\bibnamefont {Baur}}, \bibinfo {author}
  {\bibfnamefont {M.~P.}\ \bibnamefont {da~Silva}}, \ and\ \bibinfo {author}
  {\bibfnamefont {A.}~\bibnamefont {Wallraff}},\ }\href@noop {} {\bibfield
  {journal} {\bibinfo  {journal} {Nature}\ }\textbf {\bibinfo {volume} {481}},\
  \bibinfo {pages} {170} (\bibinfo {year} {2012})},\ \Eprint
  {http://arxiv.org/abs/arXiv:1108.3966} {arXiv:1108.3966} \BibitemShut
  {NoStop}%
\bibitem [{\citenamefont {Stojanovi\ifmmode~\acute{c}\else \'{c}\fi{}}\ \emph
  {et~al.}(2012)\citenamefont {Stojanovi\ifmmode~\acute{c}\else \'{c}\fi{}},
  \citenamefont {Fedorov}, \citenamefont {Wallraff},\ and\ \citenamefont
  {Bruder}}]{Stojanovic12}%
  \BibitemOpen
  \bibfield  {author} {\bibinfo {author} {\bibfnamefont {V.~M.}\ \bibnamefont
  {Stojanovi\ifmmode~\acute{c}\else \'{c}\fi{}}}, \bibinfo {author}
  {\bibfnamefont {A.}~\bibnamefont {Fedorov}}, \bibinfo {author} {\bibfnamefont
  {A.}~\bibnamefont {Wallraff}}, \ and\ \bibinfo {author} {\bibfnamefont
  {C.}~\bibnamefont {Bruder}},\ }\href {\doibase 10.1103/PhysRevB.85.054504}
  {\bibfield  {journal} {\bibinfo  {journal} {Phys. Rev. B}\ }\textbf {\bibinfo
  {volume} {85}},\ \bibinfo {pages} {054504} (\bibinfo {year}
  {2012})}\BibitemShut {NoStop}%
\bibitem [{\citenamefont {Monz}\ \emph {et~al.}(2009)\citenamefont {Monz},
  \citenamefont {Kim}, \citenamefont {H\"ansel}, \citenamefont {Riebe},
  \citenamefont {Villar}, \citenamefont {Schindler}, \citenamefont {Chwalla},
  \citenamefont {Hennrich},\ and\ \citenamefont {Blatt}}]{Monz09}%
  \BibitemOpen
  \bibfield  {author} {\bibinfo {author} {\bibfnamefont {T.}~\bibnamefont
  {Monz}}, \bibinfo {author} {\bibfnamefont {K.}~\bibnamefont {Kim}}, \bibinfo
  {author} {\bibfnamefont {W.}~\bibnamefont {H\"ansel}}, \bibinfo {author}
  {\bibfnamefont {M.}~\bibnamefont {Riebe}}, \bibinfo {author} {\bibfnamefont
  {A.~S.}\ \bibnamefont {Villar}}, \bibinfo {author} {\bibfnamefont
  {P.}~\bibnamefont {Schindler}}, \bibinfo {author} {\bibfnamefont
  {M.}~\bibnamefont {Chwalla}}, \bibinfo {author} {\bibfnamefont
  {M.}~\bibnamefont {Hennrich}}, \ and\ \bibinfo {author} {\bibfnamefont
  {R.}~\bibnamefont {Blatt}},\ }\href {\doibase 10.1103/PhysRevLett.102.040501}
  {\bibfield  {journal} {\bibinfo  {journal} {Phys. Rev. Lett.}\ }\textbf
  {\bibinfo {volume} {102}},\ \bibinfo {pages} {040501} (\bibinfo {year}
  {2009})}\BibitemShut {NoStop}%
\bibitem [{\citenamefont {Borrelli}\ \emph {et~al.}(2011)\citenamefont
  {Borrelli}, \citenamefont {Mazzola}, \citenamefont {Paternostro},\ and\
  \citenamefont {Maniscalco}}]{Borrelli11}%
  \BibitemOpen
  \bibfield  {author} {\bibinfo {author} {\bibfnamefont {M.}~\bibnamefont
  {Borrelli}}, \bibinfo {author} {\bibfnamefont {L.}~\bibnamefont {Mazzola}},
  \bibinfo {author} {\bibfnamefont {M.}~\bibnamefont {Paternostro}}, \ and\
  \bibinfo {author} {\bibfnamefont {S.}~\bibnamefont {Maniscalco}},\ }\href
  {\doibase 10.1103/PhysRevA.84.012314} {\bibfield  {journal} {\bibinfo
  {journal} {Phys. Rev. A}\ }\textbf {\bibinfo {volume} {84}},\ \bibinfo
  {pages} {012314} (\bibinfo {year} {2011})}\BibitemShut {NoStop}%
\bibitem [{\citenamefont {Ralph}\ \emph {et~al.}(2007)\citenamefont {Ralph},
  \citenamefont {Resch},\ and\ \citenamefont {Gilchrist}}]{Ralph07}%
  \BibitemOpen
  \bibfield  {author} {\bibinfo {author} {\bibfnamefont {T.~C.}\ \bibnamefont
  {Ralph}}, \bibinfo {author} {\bibfnamefont {K.~J.}\ \bibnamefont {Resch}}, \
  and\ \bibinfo {author} {\bibfnamefont {A.}~\bibnamefont {Gilchrist}},\ }\href
  {\doibase 10.1103/PhysRevA.75.022313} {\bibfield  {journal} {\bibinfo
  {journal} {Phys. Rev. A}\ }\textbf {\bibinfo {volume} {75}},\ \bibinfo
  {pages} {022313} (\bibinfo {year} {2007})}\BibitemShut {NoStop}%
\bibitem [{\citenamefont {Lanyon}\ \emph {et~al.}(2008)\citenamefont {Lanyon},
  \citenamefont {Barbieri}, \citenamefont {Almeida}, \citenamefont {Jennewein},
  \citenamefont {Ralph}, \citenamefont {Resch}, \citenamefont {Pryde},
  \citenamefont {O'Brien}, \citenamefont {Gilchrist},\ and\ \citenamefont
  {White}}]{Lanyon08}%
  \BibitemOpen
  \bibfield  {author} {\bibinfo {author} {\bibfnamefont {B.~P.}\ \bibnamefont
  {Lanyon}}, \bibinfo {author} {\bibfnamefont {M.}~\bibnamefont {Barbieri}},
  \bibinfo {author} {\bibfnamefont {M.~P.}\ \bibnamefont {Almeida}}, \bibinfo
  {author} {\bibfnamefont {T.}~\bibnamefont {Jennewein}}, \bibinfo {author}
  {\bibfnamefont {T.~C.}\ \bibnamefont {Ralph}}, \bibinfo {author}
  {\bibfnamefont {K.~J.}\ \bibnamefont {Resch}}, \bibinfo {author}
  {\bibfnamefont {G.~J.}\ \bibnamefont {Pryde}}, \bibinfo {author}
  {\bibfnamefont {J.~L.}\ \bibnamefont {O'Brien}}, \bibinfo {author}
  {\bibfnamefont {A.}~\bibnamefont {Gilchrist}}, \ and\ \bibinfo {author}
  {\bibfnamefont {A.~G.}\ \bibnamefont {White}},\ }\href@noop {} {\bibfield
  {journal} {\bibinfo  {journal} {Nature Physics}\ }\textbf {\bibinfo {volume}
  {5}},\ \bibinfo {pages} {134} (\bibinfo {year} {2008})}\BibitemShut {NoStop}%
\bibitem [{\citenamefont {Shao}\ \emph {et~al.}(2007)\citenamefont {Shao},
  \citenamefont {Zhu}, \citenamefont {Zhang}, \citenamefont {Chung},\ and\
  \citenamefont {Yeon}}]{Shao07}%
  \BibitemOpen
  \bibfield  {author} {\bibinfo {author} {\bibfnamefont {X.-Q.}\ \bibnamefont
  {Shao}}, \bibinfo {author} {\bibfnamefont {A.-D.}\ \bibnamefont {Zhu}},
  \bibinfo {author} {\bibfnamefont {S.}~\bibnamefont {Zhang}}, \bibinfo
  {author} {\bibfnamefont {J.-S.}\ \bibnamefont {Chung}}, \ and\ \bibinfo
  {author} {\bibfnamefont {K.-H.}\ \bibnamefont {Yeon}},\ }\href {\doibase
  10.1103/PhysRevA.75.034307} {\bibfield  {journal} {\bibinfo  {journal} {Phys.
  Rev. A}\ }\textbf {\bibinfo {volume} {75}},\ \bibinfo {pages} {034307}
  (\bibinfo {year} {2007})}\BibitemShut {NoStop}%
\bibitem [{\citenamefont {Vartiainen}\ \emph {et~al.}(2004)\citenamefont
  {Vartiainen}, \citenamefont {M\"ott\"onen},\ and\ \citenamefont
  {Salomaa}}]{Mikko}%
  \BibitemOpen
  \bibfield  {author} {\bibinfo {author} {\bibfnamefont {J.~J.}\ \bibnamefont
  {Vartiainen}}, \bibinfo {author} {\bibfnamefont {M.}~\bibnamefont
  {M\"ott\"onen}}, \ and\ \bibinfo {author} {\bibfnamefont {M.~M.}\
  \bibnamefont {Salomaa}},\ }\href {\doibase 10.1103/PhysRevLett.92.177902}
  {\bibfield  {journal} {\bibinfo  {journal} {Phys. Rev. Lett.}\ }\textbf
  {\bibinfo {volume} {92}},\ \bibinfo {pages} {177902} (\bibinfo {year}
  {2004})}\BibitemShut {NoStop}%
\bibitem [{\citenamefont {Shende}\ \emph {et~al.}(2006)\citenamefont {Shende},
  \citenamefont {Bullock},\ and\ \citenamefont {Markov}}]{Shende06}%
  \BibitemOpen
  \bibfield  {author} {\bibinfo {author} {\bibfnamefont {V.~V.}\ \bibnamefont
  {Shende}}, \bibinfo {author} {\bibfnamefont {S.~S.}\ \bibnamefont {Bullock}},
  \ and\ \bibinfo {author} {\bibfnamefont {I.~L.}\ \bibnamefont {Markov}},\
  }\href {\doibase 10.1109/TCAD.2005.855930} {\bibfield  {journal} {\bibinfo
  {journal} {IEEE Trans. on CAD}\ }\textbf {\bibinfo {volume} {25}},\ \bibinfo
  {pages} {1000} (\bibinfo {year} {2006})},\ \Eprint
  {http://arxiv.org/abs/arXiv:quant-ph/0406176} {arXiv:quant-ph/0406176}
  \BibitemShut {NoStop}%
\bibitem [{\citenamefont {Saeedi}\ \emph {et~al.}(2011)\citenamefont {Saeedi},
  \citenamefont {Arabzadeh}, \citenamefont {{Saheb Zamani}},\ and\
  \citenamefont {Sedighi}}]{SaeediQIC11}%
  \BibitemOpen
  \bibfield  {author} {\bibinfo {author} {\bibfnamefont {M.}~\bibnamefont
  {Saeedi}}, \bibinfo {author} {\bibfnamefont {M.}~\bibnamefont {Arabzadeh}},
  \bibinfo {author} {\bibfnamefont {M.}~\bibnamefont {{Saheb Zamani}}}, \ and\
  \bibinfo {author} {\bibfnamefont {M.}~\bibnamefont {Sedighi}},\ }\href@noop
  {} {\bibfield  {journal} {\bibinfo  {journal} {Quant. Inf. Comput.}\ }\textbf
  {\bibinfo {volume} {11}},\ \bibinfo {pages} {0262} (\bibinfo {year}
  {2011})},\ \Eprint {http://arxiv.org/abs/arXiv:1011.2159} {arXiv:1011.2159}
  \BibitemShut {NoStop}%
\bibitem [{\citenamefont {Shende}\ and\ \citenamefont
  {Markov}(2009)}]{Shende09}%
  \BibitemOpen
  \bibfield  {author} {\bibinfo {author} {\bibfnamefont {V.~V.}\ \bibnamefont
  {Shende}}\ and\ \bibinfo {author} {\bibfnamefont {I.~L.}\ \bibnamefont
  {Markov}},\ }\href@noop {} {\bibfield  {journal} {\bibinfo  {journal} {Quant.
  Inf. Comput.}\ }\textbf {\bibinfo {volume} {9}},\ \bibinfo {pages} {461}
  (\bibinfo {year} {2009})},\ \Eprint {http://arxiv.org/abs/arXiv:0803.2316}
  {arXiv:0803.2316} \BibitemShut {NoStop}%
\bibitem [{\citenamefont {Wang}\ \emph {et~al.}(2001)\citenamefont {Wang},
  \citenamefont {S\o{}rensen},\ and\ \citenamefont {M\o{}lmer}}]{Wang01}%
  \BibitemOpen
  \bibfield  {author} {\bibinfo {author} {\bibfnamefont {X.}~\bibnamefont
  {Wang}}, \bibinfo {author} {\bibfnamefont {A.}~\bibnamefont {S\o{}rensen}}, \
  and\ \bibinfo {author} {\bibfnamefont {K.}~\bibnamefont {M\o{}lmer}},\ }\href
  {\doibase 10.1103/PhysRevLett.86.3907} {\bibfield  {journal} {\bibinfo
  {journal} {Phys. Rev. Lett.}\ }\textbf {\bibinfo {volume} {86}},\ \bibinfo
  {pages} {3907} (\bibinfo {year} {2001})}\BibitemShut {NoStop}%
\bibitem [{\citenamefont {Ivanov}\ and\ \citenamefont
  {Vitanov}(2011)}]{Ivanov11}%
  \BibitemOpen
  \bibfield  {author} {\bibinfo {author} {\bibfnamefont {S.~S.}\ \bibnamefont
  {Ivanov}}\ and\ \bibinfo {author} {\bibfnamefont {N.~V.}\ \bibnamefont
  {Vitanov}},\ }\href {\doibase 10.1103/PhysRevA.84.022319} {\bibfield
  {journal} {\bibinfo  {journal} {Phys. Rev. A}\ }\textbf {\bibinfo {volume}
  {84}},\ \bibinfo {pages} {022319} (\bibinfo {year} {2011})}\BibitemShut
  {NoStop}%
\bibitem [{\citenamefont {Duan}\ \emph {et~al.}(2005)\citenamefont {Duan},
  \citenamefont {Wang},\ and\ \citenamefont {Kimble}}]{Duan05}%
  \BibitemOpen
  \bibfield  {author} {\bibinfo {author} {\bibfnamefont {L.-M.}\ \bibnamefont
  {Duan}}, \bibinfo {author} {\bibfnamefont {B.}~\bibnamefont {Wang}}, \ and\
  \bibinfo {author} {\bibfnamefont {H.~J.}\ \bibnamefont {Kimble}},\ }\href
  {\doibase 10.1103/PhysRevA.72.032333} {\bibfield  {journal} {\bibinfo
  {journal} {Phys. Rev. A}\ }\textbf {\bibinfo {volume} {72}},\ \bibinfo
  {pages} {032333} (\bibinfo {year} {2005})}\BibitemShut {NoStop}%
\bibitem [{\citenamefont {Yang}\ and\ \citenamefont {Han}(2005)}]{Yang05}%
  \BibitemOpen
  \bibfield  {author} {\bibinfo {author} {\bibfnamefont {C.-P.}\ \bibnamefont
  {Yang}}\ and\ \bibinfo {author} {\bibfnamefont {S.}~\bibnamefont {Han}},\
  }\href {\doibase 10.1103/PhysRevA.72.032311} {\bibfield  {journal} {\bibinfo
  {journal} {Phys. Rev. A}\ }\textbf {\bibinfo {volume} {72}},\ \bibinfo
  {pages} {032311} (\bibinfo {year} {2005})}\BibitemShut {NoStop}%
\bibitem [{\citenamefont {Abdollahi}\ \emph {et~al.}(2013)\citenamefont
  {Abdollahi}, \citenamefont {Saeedi},\ and\ \citenamefont
  {Pedram}}]{Abdollahi}%
  \BibitemOpen
  \bibfield  {author} {\bibinfo {author} {\bibfnamefont {A.}~\bibnamefont
  {Abdollahi}}, \bibinfo {author} {\bibfnamefont {M.}~\bibnamefont {Saeedi}}, \
  and\ \bibinfo {author} {\bibfnamefont {M.}~\bibnamefont {Pedram}},\
  }\href@noop {} {\bibfield  {journal} {\bibinfo  {journal} {Quant. Info.
  Comput.}\ } (\bibinfo {year} {2013})},\ \Eprint
  {http://arxiv.org/abs/arXiv:1302.5382} {arXiv:1302.5382} \BibitemShut
  {NoStop}%
\bibitem [{\citenamefont {Kutin}\ \emph {et~al.}()\citenamefont {Kutin},
  \citenamefont {Moulton},\ and\ \citenamefont {Smithline}}]{Kutin07}%
  \BibitemOpen
  \bibfield  {author} {\bibinfo {author} {\bibfnamefont {S.}~\bibnamefont
  {Kutin}}, \bibinfo {author} {\bibfnamefont {D.}~\bibnamefont {Moulton}}, \
  and\ \bibinfo {author} {\bibfnamefont {L.}~\bibnamefont {Smithline}},\
  }\href@noop {} {\bibfield  {journal} {\bibinfo  {journal} {Chicago J. of
  Theor. Comput. Sci.}\ }}\Eprint {http://arxiv.org/abs/arXiv:quant-ph/0701194}
  {arXiv:quant-ph/0701194} \BibitemShut {NoStop}%
\bibitem [{\citenamefont {Maslov}\ \emph {et~al.}(2008)\citenamefont {Maslov},
  \citenamefont {Dueck}, \citenamefont {Miller},\ and\ \citenamefont
  {Negrevergne}}]{MaslovTCAD}%
  \BibitemOpen
  \bibfield  {author} {\bibinfo {author} {\bibfnamefont {D.}~\bibnamefont
  {Maslov}}, \bibinfo {author} {\bibfnamefont {G.~W.}\ \bibnamefont {Dueck}},
  \bibinfo {author} {\bibfnamefont {D.~M.}\ \bibnamefont {Miller}}, \ and\
  \bibinfo {author} {\bibfnamefont {C.}~\bibnamefont {Negrevergne}},\
  }\href@noop {} {\bibfield  {journal} {\bibinfo  {journal} {IEEE Trans. on CAD
  of Integrated Circuits and Systems}\ }\textbf {\bibinfo {volume} {27}},\
  \bibinfo {pages} {436} (\bibinfo {year} {2008})},\ \Eprint
  {http://arxiv.org/abs/arXiv:quant-ph/0604001} {arXiv:quant-ph/0604001}
  \BibitemShut {NoStop}%
\bibitem [{\citenamefont {Barenco}\ \emph {et~al.}(1996)\citenamefont
  {Barenco}, \citenamefont {Ekert}, \citenamefont {Suominen},\ and\
  \citenamefont {T\"orm\"a}}]{BarencoAQFT}%
  \BibitemOpen
  \bibfield  {author} {\bibinfo {author} {\bibfnamefont {A.}~\bibnamefont
  {Barenco}}, \bibinfo {author} {\bibfnamefont {A.}~\bibnamefont {Ekert}},
  \bibinfo {author} {\bibfnamefont {K.-A.}\ \bibnamefont {Suominen}}, \ and\
  \bibinfo {author} {\bibfnamefont {P.}~\bibnamefont {T\"orm\"a}},\ }\href
  {\doibase 10.1103/PhysRevA.54.139} {\bibfield  {journal} {\bibinfo  {journal}
  {Phys. Rev. A}\ }\textbf {\bibinfo {volume} {54}},\ \bibinfo {pages} {139}
  (\bibinfo {year} {1996})}\BibitemShut {NoStop}%
\bibitem [{\citenamefont {Riebe}\ \emph {et~al.}(2006)\citenamefont {Riebe},
  \citenamefont {Kim}, \citenamefont {Schindler}, \citenamefont {Monz},
  \citenamefont {Schmidt}, \citenamefont {K\"orber}, \citenamefont {H\"ansel},
  \citenamefont {H\"affner}, \citenamefont {Roos},\ and\ \citenamefont
  {Blatt}}]{Riebe}%
  \BibitemOpen
  \bibfield  {author} {\bibinfo {author} {\bibfnamefont {M.}~\bibnamefont
  {Riebe}}, \bibinfo {author} {\bibfnamefont {K.}~\bibnamefont {Kim}}, \bibinfo
  {author} {\bibfnamefont {P.}~\bibnamefont {Schindler}}, \bibinfo {author}
  {\bibfnamefont {T.}~\bibnamefont {Monz}}, \bibinfo {author} {\bibfnamefont
  {P.~O.}\ \bibnamefont {Schmidt}}, \bibinfo {author} {\bibfnamefont {T.~K.}\
  \bibnamefont {K\"orber}}, \bibinfo {author} {\bibfnamefont {W.}~\bibnamefont
  {H\"ansel}}, \bibinfo {author} {\bibfnamefont {H.}~\bibnamefont {H\"affner}},
  \bibinfo {author} {\bibfnamefont {C.~F.}\ \bibnamefont {Roos}}, \ and\
  \bibinfo {author} {\bibfnamefont {R.}~\bibnamefont {Blatt}},\ }\href
  {\doibase 10.1103/PhysRevLett.97.220407} {\bibfield  {journal} {\bibinfo
  {journal} {Phys. Rev. Lett.}\ }\textbf {\bibinfo {volume} {97}},\ \bibinfo
  {pages} {220407} (\bibinfo {year} {2006})}\BibitemShut {NoStop}%
\bibitem [{\citenamefont {Yang}\ \emph {et~al.}(2010)\citenamefont {Yang},
  \citenamefont {Liu},\ and\ \citenamefont {Nori}}]{Yang}%
  \BibitemOpen
  \bibfield  {author} {\bibinfo {author} {\bibfnamefont {C.-P.}\ \bibnamefont
  {Yang}}, \bibinfo {author} {\bibfnamefont {Y.-x.}\ \bibnamefont {Liu}}, \
  and\ \bibinfo {author} {\bibfnamefont {F.}~\bibnamefont {Nori}},\ }\href
  {\doibase 10.1103/PhysRevA.81.062323} {\bibfield  {journal} {\bibinfo
  {journal} {Phys. Rev. A}\ }\textbf {\bibinfo {volume} {81}},\ \bibinfo
  {pages} {062323} (\bibinfo {year} {2010})}\BibitemShut {NoStop}%
\end{thebibliography}

%

\section*{Acknowledgements}
Authors were supported by the Intelligence Advanced
Research Projects Activity (IARPA) via Department
of Interior National Business Center contract number
D11PC20165. The U.S. Government is authorized to reproduce
and distribute reprints for Governmental purposes
notwithstanding any copyright annotation thereon. The views
and conclusions contained herein are those of the authors and
should not be interpreted as necessarily representing the official
policies or endorsements, either expressed or implied, of
IARPA, DoI/NBC, or the U.S. Government.

\end{document}